\documentclass[12pt]{article}
\def \d{\partial}

\def \r{{\bf r}}
\def \be{\begin{equation}}
\def \ee{\end{equation}}

\def \1{{\bf 1}}

\def \gg{\gamma}
\def \dd{\delta}
\def \eps{\epsilon}

\def \ll{\lambda}
\def \om{\omega}
\def \ss{\sigma}
\def \bra{{\bar a}}
\def \brb{{\bar b}}
\def \brc{{\bar c}}

\def \bre{{\bar e}}

\def \brg{{\bar g}}
\def \brh{{\bar h}}
\def \brj{{\bar j}}

\def \bom{{\bar \om}}

\def \BA{{\bar A}}

\def \BC{{\bar C}}
\def \BD{{\bar D}}
\def \BR{{\bar R}}
\def \GG{\Gamma}
\def \LL{\Lambda}
\def \BGG{{\bar \GG}}
\def \BT{{\bar T}}

\def \Tr{\mbox{Tr}}
\def \fr{\frac}

\def\bec{\begin{center}}
\def\enc{\end{center}}

\def\bef{\begin{figure}[htb]}
\def\befh{\begin{figure}[!h!]}
\def\enf{\end{figure}}
\usepackage{epsfig,latexsym}
\newfont{\st}{cmr7}

\begin{document}


\title{ Bi-metric Gravity and ``Dark Matter''}

\author{ I.T.~Drummond,\\
DAMTP\\ University of Cambridge\\ Silver Street\\ Cambridge, CB3~9EW, UK}
\maketitle

\begin{abstract}
We present a bi-metric theory of gravity containing a length scale
of galactic size. For distances less than this scale the theory
satisfies the standard tests of General Relativity. For distances  
greater than this scale the theory yields an effective gravitational 
constant much larger than the locally observed value of Newton's constant. 
The transition from one regime to the other through the galactic scale can 
explain the observed rotation curves of galaxies and hence the effects 
normally attributed to the presence of dark matter. Phenomena on an 
extragalactic scale such as galactic clusters and the expansion  of 
the universe are controlled by the enhanced gravitational coupling. This 
provides an explanation of the missing matter normally invoked to account
for the observed value of Hubble's constant in relation to observed matter.
\end{abstract}
\vfill
DAMTP-00-83
\pagebreak

\section{Introduction}

The presence of dark matter in the universe has been invoked to explain
a number of phenomena, the rotation curves of galaxies\cite{RUB1}-\cite{ALB1} 
the apparent mass of galactic clusters\cite{ZWI1,CAR} and the observed expansion 
of the universe. 
Many attempts to detect this conjectured matter are under way but at the 
moment there is no direct observation of it. In this paper we present a 
model with a modified version of gravity that can accommodate these points 
in a relatively straightforward way. It does so without the necessity of 
postulating dark matter. The theory, which is geometrical in character, 
is perfectly covariant, locally Lorentz invariant and satisfies the Equivalence 
Principle. It also satisfies all the tests of General Relativity on the scales 
at which they have been carried out.

The theory is a bi-metric theory. Such theories have a long history \cite{R1,R2,R3}
and have recently been used as a way of realising Variable Speed of Light (VSL) 
theories \cite{AM}-\cite{BAS}. 
An earlier incomplete version of the theory presented here was motivated
in this way \cite{ITD}. Bi-metric theories ran into difficulties when, as then formulated, 
they appeared to be inconsistent with the tests of General Relativity \cite{CW1,CW2}.
However the theory proposed here does meet all those tests \cite{CW1,CW2,HUL,HULTAY,HUL,TAY}. 
Of course VSL is built into bi-metric theories. This will be important in applying 
the model to early universe studies.  However in this paper we concentrate on those 
implications of our theory for the gravitational interaction of matter that provide 
an explanation of the current state of the universe without recourse to dark matter.

Our proposal is to introduce into the space-time manifold, two vierbein bundles.
Each bundle supports its own metric.
One is associated with matter and the other with (underlying) gravity. The matter vierbein
can be strained and scaled relative to the gravitational vierbein.  The dynamics of the 
theory includes this straining and scaling as
dynamical degrees of freedom. The justification for introducing these new
effects is ultimately in the results that emerge. However if dark matter is not present
some modification of gravity or dynamics is essential. We choose to modify gravity 
in a way that permits the introduction of a galactic scale, something that is impossible
with standard General Relativity.

The equations of motion are obtained from an action comprising three parts, a 
gravitational term $I_G$, of the standard curvature form, a matter term $I_M$, 
of the  standard form based on the matter metric, and  a linking action $I_L$, 
that depends on the variables that determine the relationship between the two 
vierbein bundles. The full action, $I$, is the sum of all three terms,
\be
I=I_G+I_L+I_M~~.
\ee
Each term has its own gravitational coupling constant with the dimensions of Newton's
constant, $G_N$~. In developing the theory we show how these different constants
are related to one another and to Newton's constant.

\section{General Structure}

The geometrical character of the theory is clearly revealed by its formulation in 
the vierbein formalism. The local Lorentz invariance in both the gravitational and
matter vierbein frame is explicit throughout as a gauge invariance. 
We introduce a vierbein bundle appropriate to gravity, $\{e_{\mu a}\}$,
with the associated metric
\be
g_{\mu\nu}=e_{\mu a}e_{\nu}^{~~a}~~,
\ee
where raising and lowering of $a$-indices is carried out with the standard
Lorentz metric $\eta_{ab}=\{1,-1,-1,-1\}$~. The inverse vierbein is $\{e^{a\mu}\}$
so that
\be
e_{\mu a}e^{a\nu}=\dd^{\nu}_{\mu}~~,~~~~~~~~e^{a\mu}e_{\mu b}=\dd^{a}_{b}~~,
\ee
and
\be
g^{\mu\nu}=e^{a\mu}e_a^{~~\nu}~~.
\ee
The vierbein associated with matter is $\{\bre_{\mu\bra}\}$ and the raising and lowering
of $\bra$-indices is by means of the Lorentz metric, $\eta_{\bra\brb}=\{1,-1,-1,-1\}$~.
The associated metric is
\be
\brg_{\mu\nu}=\bre_{\mu\bra}\bre_{\nu}^{~~\bra}~~.
\ee
The two vierbein bundles are related by a local linear transformation
\be
\bre_{\mu\bra}=e_{\mu a}M^a_{~~\bra}e^\phi~~,
\ee
where the matrix $M$ is an element of $SL(4,R)$ and the local scaling is introduced through
the factor $e^\phi$~. The determinants associated with the volume elements of the
bundles are $J$ and $\bar J$ where
\be
J=\det\{e_\mu^{~~a}\}~~,~~~~~~~~\hbox{and}~~~~~~~~ \bar J=\det\{\bre_\mu^{~~\bra}\}~~.
\ee
We have therefore ${\bar J}=Je^{4\phi}$~.
We denote the inverse matrix by $M^{\bra}_{~~a}$ so that
\be
M^a_{~~\bra}M^{\bra}_{~~b}=\dd^a_b~~,~~~~~~~~M^{\bra}_{~~a}M^a_{~~\brb}=\dd^{\bra}_{\brb}~~.
\ee

The vierbein connections and associated coordinate connections are defined so that
the vierbeins are covariantly constant in the appropriate way.
\be
D_\mu e_{\nu a}=\d_\mu e_{\nu a}+\om_{\mu a}^{~~~b}e_{\nu b}-\GG^\ll_{\mu\nu}e_{\ll a}=0~~,
\ee
and
\be
\BD_\mu\bre_{\nu\bra}
=\d_\mu\bre_{\nu\bra}+\bom_{\mu\bra}^{~~~\brb}\bre_{\nu\brb}-\BGG^\ll_{\mu\nu}\bre_{\ll \bra}=0~~.
\ee
The requirement that $\eta_{a b}$ and $\eta_{\bra\brb}$ be covariantly constant 
implies that $\om_{\mu ab}=-\om_{\mu ba}$ and $\bom_{\mu\bra\brb}=-\bom_{\mu\brb\bra}$~.

It is convenient to define a covariant derivative of $M$ that includes both the right and left 
vierbein connections,
\be
D_\mu M^a_{~~\bra}=\d_\mu M^a_{~~\bra}+\om^{~a}_{\mu~~b}M^b_{~~\bra}
                     -M^a_{~~\brb}~\bom_{\mu~~\bra}^{~\brb}~~.
\ee
However further differentiation requires the coordinate connection, 
$\GG^\ll_{\mu\nu}$~. The covariant deriviative $\BD_\mu$ can be defined and 
extended in a similar way. Its effect on $M$ is the same as that of $D_\mu$ but
a second differentiation must use the coordinate connection $\BGG^\ll_{\mu\nu}$~. 

Gravitational curvature tensors appropriate to each of the bundles are defined so that
\be
\left[D^L_\mu,D^L_\nu\right]V_a=R_{ab\mu\nu}V^b~~,
\ee
and
\be
\left[D^R_\mu,D^R_\nu\right]V_\bra=\BR_{\bra\brb\mu\nu}V^\brb~~,
\ee
where $D^L_\mu$ includes the left vierbein connection field, $\om_{\mu ab}$ but not
the coordinate connection, $\GG^\ll_{\mu\nu}$~. Similarly $D^R_\mu$ includes the only the
right vierbein connection, $\bom_{\mu \bra\brb}$~. We have then
\be
R_{ab\mu\nu}=\d_\mu\om_{\nu ab}-\d_\nu\om_{\mu ab}
 +\om_{\mu a}^{~~~c}~\om_{\nu cb}-\om_{\nu a}^{~~~c}~\om_{\mu cb}~~,
\label{CURV1}
\ee
with a similar definition for $\BR_{\bra\brb\mu\nu}$~. It follows that
\be
\left[D_\mu,D_\nu\right]M^a_{~~\bra}
          =R^a_{~~b\mu\nu}M^b_{~~\bra}-M^a_{~~\brb}\BR^\brb_{~~\bra\mu\nu}
       -2C^\ll_{\mu\nu}D_\ll M^a_{~~\bra}~~,
\label{CURV2}
\ee
where
\be
C^\ll_{\mu\nu}=\fr{1}{2}\left(\GG^\ll_{\mu\nu}-\GG^\ll_{\nu\mu}\right)~~.
\ee
The quantity $C^\ll_{\mu\nu}$ is the torsion tensor in the coordinate basis. It
is not in general zero.

\section{Gravitational Action}
The gravitational action has the standard form
\be
I_G=-\fr{1}{16\pi G}\int d^4xJR~~,
\ee
where
\be
R=e^{a\mu}e^{b\nu}R_{ab\mu\nu}~~,
\ee
and $G$ is a coupling with the dimensions of Newton's constant, $G_N$~.
The vierbeins and the connections are treated as independent variables.
If we vary the former then
\be
\dd I_G=\fr{1}{8\pi G}\int d^4xJ\dd e_{\ss c}\left(e^{c\mu}R^\ss_{~~\mu}-\fr{1}{2}e^{c\ss}R\right)~~.
\ee
The variation of the vierbein connection yields
\be
\dd I_G=-\fr{1}{16\pi G}\int d^4xJe^{a\mu}e^{b\nu}\left(D^L_\mu\dd\om_{\nu ab}
                                             -D^L_\nu\dd\om_{\mu ab}\right)~~.
\ee
Switching to the full covariant derivative we get
\be
\dd I_G=-\fr{1}{16\pi G}\int d^4xJe^{a\mu}e^{b\nu}
 \left(D_\mu\dd\om_{\nu ab}-D_\nu\dd\om_{\mu ab}+2C^\ll_{\mu\nu}\dd\om_{\ll ab}\right)~~.
\ee
Using the result
\be
JD_\mu V^\mu=\d_\mu(JV^\mu)+2JC^\ll_{\ll\mu}V^\mu~~.
\ee
we can integrate by parts and obtain finally
\be
\dd I_G=-\fr{1}{8\pi G}\int d^4xJe^{a\mu}e^{b\nu}\left(C^\ll_{\ll\mu}\dd^\ss_\nu
    -C^\ll_{\ll\nu}\dd^\ss_\mu+C^\ss_{\mu\nu}\right)\dd\om_{\ss ab}~~.
\ee
If there is no other interaction in the theory we can deduce
from the vanishing of these variations that $C^\ll_{\mu\nu}=0$, and 
$$
R^\ss_{~~\mu}-\fr{1}{2}\dd^\ss_\mu R=0~~,
$$
the standard equations for matterless gravity.  

\section{Linking Action}

Proposals for constructing an action for the linear transformation relating the
bundles have involved parametrizing it in terms of a vector or scalar field 
\cite{M1,M2,CM1,CM2,BAS}. Our proposal treats all the degrees of freedom inherent 
in the transformation. This is crucial for the structure of the theory. 

Because scaling commutes with the other elements of the group of linear transformations, 
it can be treated separately. The  Lagrangian for $\phi$ can be chosen to be proportional 
to $g^{\mu\nu}\d_\mu\phi\d_\nu\phi$~. The remaining degrees of freedom are represented
by the matrix $M$ which lies in the non-linear manifold $SL(4,R)$~. A natural way of
constructung a Lagrangian for such a theory is to invoke the mechanism of the non-linear
sigma model and express it as a quadratic form in the derivatives $(D_\mu M)M^{-1}$
with the appropriate structure. The use of the covariant derivative guarantees the
local Lorentz gauge invariance relative to both bundles and the presence of $M^{-1}$ 
guarantees that the derivative is a proper element of the tangent space to the 
manifold $SL(4,R)$~. We take as our action 

\be
I_L=\fr{1}{16\pi F}\int d^4xJg^{\mu\nu}\Tr(j_\mu j_\nu)
          +\fr{1}{16\pi F'}\int d^4xJg^{\mu\nu}(\d_\mu\phi \d_\nu\phi)~~,
\ee
where $F$ and $F'$ are new gravitational constants with the same physical dimensions as $G$~. 
The matrix valued current $j_\mu$ is given by
\be
j_\mu=(D_\mu M)M^{-1}~~,
\ee
or more explicitly
\be
j_\mu^{ab}=(D_\mu M^a_{~~\brb})M^{\brb b}~~.
\ee

It is also convenient to define an alternative version of the current,
appropriate to the barred vierbein bundle, $\brj_\mu^{\bra\brb}$, as
\be
\brj_\mu^{\bra\brb}=\left(M^{-1}D_\mu M\right)^{\bra\brb}=M^{\bra}_{~~a}D_\mu M^{a\brb}~~.
\ee

We also include in $I_L$ ``mass'' terms of the form 
\be
-\fr{1}{16\pi F}\int d^4xJ\fr{m^2}{4}\left(M^a_{~~\bra}M_a^{~~\bra}
                                      +M_{\bra}^{~~a}M^{\bra}_{~~a}-\gamma\right)
-\fr{1}{16\pi F'}\int d^4xJm^2\phi^2~~.
\ee
For simplicity we have chosen the mass parameter $m$ to be the same in both these 
these additional terms. For the choice $\gamma=8$, the action vanishes when $M$ represents
a Lorentz transformation. Departures from this value introduce a cosmological
constant term in the action. By construction these additional terms 
clearly maintain local Lorentz invariance in both vierbein bundles.
The mass terms are crucial for the effectiveness of the theory because we identify
the galactic distance scale with $m^{-1}$~. For effects
on a scale much less than this therefore $m$ will be viewed as a small parameter
that can be neglected in certain circumstances.

We treat the vierbein, which enters through $g^{\mu\nu}$, the matrix, $M$, and
the connections $\om_{\mu ab}$ and $\bom_{\mu\bra\brb}$ as independent variables.
The result for the linking action from the vierbein variation is,
\begin{eqnarray}
\dd I_L=-\fr{1}{8\pi F}\int d^4xJ\dd e_{\ss c}
     \left(e^{c\nu}g^{\ss\mu}-\fr{1}{2}e^{c\ss}g^{\mu\nu}\right)
     \Tr(j_\mu j_\nu)~~~~~~~~~~~~~~~~~~~~~~~~~~~~~\nonumber\\
  -\fr{1}{8\pi F}\int d^4xJ\dd e_{\ss c}e^{c \ss}\fr{m^2}{8}
      \left(M^a_{~~\bra}M_a^{~~\bra}+M^\bra_{~~a}M_\bra^{~~a}-\gamma\right)~~~~\nonumber\\
   -\fr{1}{8\pi F'}\int d^4xJ\dd e_{\ss c}
           \left(e^{c\nu}g^{\ss\mu}-\fr{1}{2}e^{c\ss}g^{\mu\nu}\right)
               \d_\mu\phi\d_\nu\phi~~~~\nonumber\\
-\fr{1}{8\pi F'}\int d^4xJ\dd e_{\ss c}e^{c \ss}\fr{m^2}{2}\phi^2~~.~~~~~~~~~~~~~~~~
\end{eqnarray}
From the left vierbein connection we have
\be
\dd I_L=\fr{1}{8\pi F}\int d^4xJ\dd\om_{\mu ab}j^{\mu ba}~~,
\ee
and from the right vierbein connection
\be
\dd I_L=-\fr{1}{8\pi F}\int d^4xJ\dd\bom_{\mu\bra\brb}\brj^{\mu\brb\bra}~~.
\ee
On varying the matrix $M$ we obtain
\be
\dd I_L=-\fr{1}{8\pi F}\int d^4xJ
   (\dd MM^{-1})^{ab}\left[D_\mu j^{\mu}_{ba}-2C^\ll_{\ll\mu}j^{\mu}_{ba}
               +\fr{m^2}{4}\left(M_a^{~~\brc} M_{b\brc}-M^\brc_{~~a}M_{\brc b}\right)\right]~~,
\ee
where the square brackets $[\cdots ]$ indicate the traceless version of the quantity
contained within them. The quantity $\dd MM^{-1}$, being an arbitrary element of the 
$SL(4,R)$ Lie algebra, is sufficiently general to identify the other factor in the integrand. 
Finally, on varying $\phi$, we have
\be
\dd I_L
  =-\fr{1}{8\pi F'}\int d^4xJ\dd\phi\left(g^{\mu\nu}\left(D_\nu\d_\mu\phi
                                   -2C^\ll_{\ll\nu}\d_\mu\phi\right)+m^2\phi\right)~~.
\ee

\section{Matter Action}

We assume that matter is propagated in the vierbein background $\{\bre_{\mu\bra}\}$~.
This seems a consistent approach since it implies that matter behaves in a conventional way
in relation to the gravitational field it experiences. In particular the Equivalence
Principle is satisfied. However the theory does change the relationship of this observed 
gravitational field to the distribution of matter density. We have 
\be
\dd I_M=-\fr{1}{2}\int d^4x{\bar J}\dd\brg_{\mu\nu}\BT^{\mu\nu}~~,
\ee
where $\BT^{\mu\nu}=\BT^{\nu\mu}$ is the symmetric energy momentum tensor for matter.
Since
\be
\brg_{\mu\nu}=\bre_{\mu\bra}\bre_\nu^{~~\bra}~~,
\ee
it follows that
\be
\dd I_M=-\int d^4x{\bar J}\dd\bre_{\mu\bra}\bre_\nu^{~~\bra}\BT^{\mu\nu}~~.
\ee
However 
\be
\dd\bre_{\mu\bra}=\dd e_{\mu a}M^a_{~~\bra}e^\phi+e_{\mu a}\dd M^a_{~~\bra}e^\phi
                                         +\dd\phi\bre_{\mu\bra}~~,
\ee
so the variation of the matter action takes the form
\be
\dd I_M=-\int d^4xJe^{4\phi}\left(\dd e_{\ss c}T^{\ss c}+\Tr(\dd MM^{-1}U)
         +\dd\phi\BT\right)~~.
\ee
where $\BT=\brg_{\mu\nu}\BT^{\mu\nu}$,
\be
U^{ba}=e_\mu^{~~a}M^b_{~~\bra}\bre_\nu^{~~\bra}\BT^{\mu\nu}-\fr{1}{4}\eta^{ab}\BT~~,
\ee
and
\be
T^{\ss c}=M^c_{~~\bra}\bre_\nu^{~~\bra}\BT^{\ss\nu}~~,
\ee
\be
T^\ss_{~~\ll}=e_{\ll c}T^{\ss c}=\bre_{\ll\bra}\bre_\nu^{~~\bra}\BT^{\ss\nu}=\BT^{\ss\nu}\brg_{\ll\nu}~~.
\ee
If we adopt the convention that barred quantities, that is those appropriate to
the gravitational background of the matter, have spatial indices raised and lowered
with the barred metric we can define
\be
\BT^\ss_{~~\ll}=\BT^{\ss\nu}\brg_{\ll\nu}~~.
\ee
Hence we get the simple seeming result
\be
T^\ss_{~~\ll}=\BT^\ss_{~~\ll}~~.
\label{MTR1}
\ee
However it is important to recall that
\be
T^{\ss\tau}=T^\ss_{~~\ll}g^{\ll\tau}\ne\BT^{\ss\tau}~~.
\ee
In fact $T^{\ss\tau}$ is not necessecarily symmetric. This does not cause any difficulty.
Note that $T=T^\mu_{~~\mu}=\BT$~.

\section{Equations of Motion}

We obtain the equations of motion by requiring that the variation
of the total action is zero. The result is
\begin{eqnarray}
\fr{1}{8\pi G}\left(R^\ss_{~~\rho}-\fr{1}{2}\dd^\ss_\rho R\right)
 -\fr{1}{8\pi F}\left(\Tr(j^\ss j_\mu)-\fr{1}{2}\dd^\ss_\mu\Tr(j^\ll j_\ll)\right)&&\nonumber\\
-\fr{1}{8\pi F}\fr{m^2}{8}\dd^\ss_\ll\left(M^a_{~~\bra}M_a^{~~\bra}+M^\bra_{~~a}M_\bra^{~~a}
                                                                  -\gamma\right)&&\nonumber\\
-\fr{1}{8\pi F'}\left(g^{\ss\nu}\d_\nu\phi \d_\mu\phi)
                 -\fr{1}{2}\dd^\ss_\mu(g^{\ll\nu}\d_\ll\phi \d_\nu\phi)\right)
     -\fr{1}{8\pi F'}\fr{m^2}{2}\dd^\ss_\ll\phi^2&=&e^{4\phi}\BT^\ss_{~~\rho}~~.
\label{EQM1}
\end{eqnarray}
Eq(\ref{EQM1}) is a generalization of the standard equation of General Relativity.
The new straining degrees of freedom are controlled by the following equation,
\be
\fr{1}{8\pi F}\left(D_\mu j^\mu_{ba}-2C^\ll_{\ll\mu}j^\mu_{ba}
       +\fr{m^2}{4}\left[M_a^{~~\brc}M_{b\brc}-M^\brc_{~~a}M_{\brc b}\right]\right)
                                 +e^{4\phi}U^{ba}=0~~.
\label{EQM2}
\ee
The generalization of the ``no torsion'' rule in General Relativity is represented by
the next two equations,
\be
j^{\ss[b,a]}=\fr{F}{G}e^{a\mu}e^{b\nu}
         \left(C^\ll_{\ll\mu}\dd^\ss_\nu-C^\ll_{\ll\nu}\dd^\ss_\mu+C^\ss_{\mu\nu}\right)~~.
\label{EQM3}
\ee
\be
\brj_\mu^{[\brb,\bra]}=0~~.
\label{EQM4}
\ee
Finally the local expansion is controlled by 
\be
\fr{1}{8\pi F'}\left(g^{\mu\nu}\left(D_\nu\d_\mu\phi-2C^\ll_{\ll\nu}\d_\mu\phi\right)+m^2\phi\right)
                          +e^{4\phi}\BT=0~~.
\label{EQM5}
\ee

Eq(\ref{EQM1}) implies that
\begin{eqnarray}
\fr{1}{8\pi G}R^\ss_{~~\mu}-\fr{1}{8\pi F}\Tr(j^\ss j_\mu)
               -\fr{1}{8\pi F'}g^{\ss\nu}\d_\nu\phi\d_\mu\phi~~~~~~~~~~~~~~~~~~~~~~~~\nonumber\\
+\fr{1}{8\pi F}\fr{m^2}{8}\dd^\ss_\ll\left(M^a_{~~\bra}M_a^{~~\bra}+M^\bra_{~~a}M_\bra^{~~a}
                                     -\gamma\right)~~~~~~~~~~~~~~~~\nonumber\\
~~~~~~~~~~~~~~~~~~~~~~~~~~~~~~~~
+\fr{1}{8\pi F'}\fr{m^2}{2}\dd^\ss_\ll\phi^2
                  =e^{4\phi}\left(\BT^\ss_{~~\mu}-\fr{1}{2}\dd^\ss_\mu \BT\right)~~.
\label{EQM1a}
\end{eqnarray}
Although rather complex these equations are surprisingly susceptible of analysis as
we show below.

\section{Bianchi Identity}

Just as in standard General Relativity it is necessary to check that the
theory satisfies the integrability conditions associated with the Bianchi identity.
In the presence of torsion this is changed to the following 
\be
R_{\ll\tau\mu\nu;\ss}+R_{\ll\tau\nu\ss;\mu}+R_{\ll\tau\ss\mu;\nu}=
 -2\left(R_{\ll\tau\rho\mu}C^\rho_{\nu\ss}+R_{\ll\tau\rho\nu}C^\rho_{\ss\mu}
+R_{\ll\tau\rho\ss}C^\rho_{\mu\nu}\right)~~,
\label{BI1}
\ee
where $;\mu$ indicates the covariant derivative $D_\mu$~. In the contracted version it becomes
\be
\left(R^\mu_{~~\ss}-\fr{1}{2}\dd^\mu_\ss R\right)_{;\mu}
=R^{\mu\nu}_{~~~\rho\ss}C^\rho_{\mu\nu}+2R^\mu_{~~\rho}C^\rho_{\ss\mu}~~.
\label{BI2}
\ee
If we take the covariant divergence of the left side of eq(\ref{EQM1}) and make use of the
equations of motion together with the identity
\be
j_\mu^{ab}=e^{a\nu}\left(\BGG^\ll_{\mu\nu}-\GG^\ll_{\mu\nu}\right)e_\ll^{~~b}-\eta^{ab}\d_\mu\phi~~,
\ee
we obtain the result
\be
\BD_\ss\BT^\ss_{~~\ll}=2\BC^\ss_{\ss\ll}\BT-2\BC^\tau_{\ss\ll}\BT^\ss_{~~\tau}~~.
\ee
It is easily checked that this is equivalent to the standard conservation law
\be
D^{(\brg)}_\ss\BT^\ss_{~~\ll}=0~~,
\ee
where $D^{(\brg)}_\ss$ is the covariant derivative formed from the metric connection
arising from $\brg_{\mu\nu}$~.

\section{Connection Structure}

In standard General Relativity the assumption that matter couples to gravity
only through the metric means that torsion plays no role in the theory.
In the theory presented here we make the same assumption about matter. However
because of the extra complexity of the theoretical structure, torsion 
is in general not zero. Nevertheless the equations of motion do permit the connection
structure to be elucidated in a straightforward way.

The starting point of the analysis is eq(\ref{EQM4}) which reveals that the 
antisymmetric part of $\brj_{\mu\bra\brb}$ vanishes. It allows us to express
$\bom_{\mu\bra\brb}$ in terms of the other dynamical variables and hence
eliminate it from the equations of motion.

We now express $j_{\mu ab}$ in terms of the symmetric part of $\brj_{\mu\bra\brb}$
as follows
\be
j_{\mu ab}=M_a^{~~\bra}\brj_{\mu\bra\brb}M^\brb_{~~b}
                  =M_a^{~~\bra}\brj_{\mu\{\bra,\brb\}}M^\brb_{~~b}~~.
\ee
We can write this more explicitly in the form
\be
j_{\mu ab}=\fr{1}{2}M_a^{~~\bra}\left(M_\bra^{~~c}\d_\mu M_{c\brb}
                    +M_\bra^{~~c}\om_{\mu c}^{~~~d}M_{d\brb}
                   +M_\brb^{~~c}\d_\mu M_{c\bra}
                    +M_\brb^{~~c}\om_{\mu c}^{~~~d}M_{d\bra}\right)M^\brb_{~~b}~~,
\label{CON1}
\ee
which shows that $j_{\mu ab}$ is a linear function of $\om_{\mu ab}$~.

The torsion tensor is also a linear function of $\om_{\mu ab}$~.
The covariant constancy of $e_{\mu a}$ implies that
\be
\GG^\ll_{\mu\nu}=e^{a\ll}\left(\d_\mu e_{\nu a}+\om_{\mu a}^{~~~b}e_{\nu b}\right)~~.
\ee
Hence 
\be
C^\ll_{\mu\nu}=\fr{1}{2}e^{a\ll}\left(\d_\mu e_{\nu a}+\om_{\mu a}^{~~~b}e_{\nu b}
                                 -\d_\nu e_{\mu a}-\om_{\nu a}^{~~~b}e_{\mu b}\right)~~.
\ee
If we define
\be
C_{\ll\mu\nu}=g_{\ll\ss}C^\ss_{\mu\nu}~~,
\ee
and 
\be
\om_{\mu\ll\nu}=e_\ll^{~~a}e_\nu^{~~b}\om_{\mu ab}~~,
\ee
then we obtain $\om_{\mu ab}$ as a linear function of the torsion,
\be
\om_{\mu\ll\nu}=C_{\ll\mu\nu}-C_{\nu\mu\ll}+C_{\mu\ll\nu}+{\hat{\om}}_{\mu\ll\nu}~~,
\ee
where ${\hat{\om}}_{\mu\ll\nu}$ is the metric version of $\om_{\mu\ll\nu}$ and is given by
\be
{\hat{\om}}_{\mu\ll\nu}=\fr{1}{2}\left(e_\nu^{~~a}\d_\mu e_{\ll a}-e_\ll^{~~a}\d_\mu e_{\nu a}
                 +\d_\nu g_{\ll\mu}-\d_\ll g_{\nu\mu}\right)~~.
\label{VBCON1}
\ee

We note that eq(\ref{EQM3}) relates the torsion tensor $C^\ll_{\mu\nu}$ linearly to 
$j_{\mu ab}$~.
\be
C_{\ss\mu\nu}+C^\ll_{\ll\mu}g_{\nu\ss}-C^\ll_{\ll\nu}g_{\mu\ss}=-X_{\ss\mu\nu}~~,
\ee
where
\be
X_{\ss\mu\nu}=\fr{G}{F}j_{\ss[\mu,\nu]}~~,
\ee
and
\be
j_{\ss\mu\nu}=e_\mu^{~~a}e_\nu^{~~b}j_{\ss ab}~~.
\ee
It follows that
\be
C_{\ss\mu\nu}=-X_{\ss\mu\nu}-\fr{1}{2}X^\ll_{\ll\mu}g_{\nu\ss}
                                   +\fr{1}{2}X^\ll_{\ll\nu}g_{\mu\ss}~~.
\ee
Finally we have the linear relation
\be
\om_{\mu\ll\nu}=-X_{\ll\mu\nu}+X_{\nu\mu\ll}-X_{\mu\ll\nu}-X^\tau_{\tau\ll}g_{\mu\nu}
        +X^\tau_{\tau\nu}g_{\mu\ll}+{\hat{\om}}_{\mu\ll\nu}~~,
\label{CON2}
\ee
that determines $\om_{\mu ab}$ in terms of the other dynamical variables.

For future reference and to show that the above equation can take a simple form
in special cases we compute $j_{\ll ab}$ when the matrix $M$ takes a diagonal form,
namely
\be
M^a_{~~\bra}=\LL_a\dd^a_\bra~~~~~~~~\mbox{and}~~~~~~~~M^{\bra}_{~~a}=\LL^{-1}_a\dd^\bra_a~~.
\ee
We enforce an obvious correspondence between the values of the symbols $a$ and $\bra$
to give meaning to the $\dd$~ symbols. 
We can compute $j_{\mu ab}$ from eq(\ref{CON1}) to obtain
\be
j_{\ll ab}=\fr{\d_\ll\LL_a}{\LL_a}\eta_{ab}
                +\fr{1}{2}\om_{\ll ab}\left(1-\fr{\LL^2_a}{\LL^2_b}\right)~~,
\label{STRCUR}
\ee
with the simple result
\be
j_{\ll[a,b]}=\fr{1}{4}\om_{\ll ab}\left(2-\fr{\LL^2_a}{\LL^2_b}-\fr{\LL^2_b}{\LL^2_a}\right)~~.
\label{VBCON2}
\ee
Used in conjunction with eq(\ref{CON2}) this leads to an easy evaluation of the 
vierbein connection.

\section{Weak Field Limit}
In order to apply it to planetary motion, galaxies and galactic clusters
it is appropriate to examine the theory in the limit of weak gravitational fields.
Since we are considering relatively local objects in an
effectively flat background we will choose the parameter $\gamma=8$ in order to 
set the cosmological constant to zero.

The weak field limit has the form
\be
e_{\mu a}=e^{(0)}_{\mu a}+h_{\mu a}~~,
\ee
where
\be
e^{(0)}_{~~\mu a}e^{(0)~a}_{~~\nu}=\eta_{\mu\nu}~~,
\ee
Similarly we can set
\be
\bre_{\mu \bra}=\bre^{(0)}_{\mu \bra}+\brh_{\mu \bra}~~,
\ee
where
\be
\bre^{(0)}_{~~\mu \bra}\bre^{(0)~\bra}_{~~\nu}=\eta_{\mu\nu}~~.
\ee
The connections $\om_{\mu ab}$, $\bom_{\mu\bra\brb}$ and the scaling field $\phi$
are first order quantities.
The matrix $M$ has the form
\be
M^a_{~~\bra}=M^{(0)a}_{~~~~~\bra}+m^a_{~~\bra}~~,
\ee
where $M^{(0)}$ is a constant Lorentz transformation. We have then
\be
M^{(0)~\bra}_{~~~~~~a}=M^{(0)~~\bra}_{~~~a}~~.
\ee
We can use $M^{(0)}$ and its inverse and $e^{(0)}$ and $\bre^{(0)}$ to convert 
superfixes and suffixes between the various bases. For example we have
\be
m^a_{~~b}=m^a_{~~\bra}M^{(0)~\bra}_{~~~~~~b}~~.
\ee
The requirement that $\det M=1$ implies that
\be
m^a_{~~a}=m^\bra_{~~\bra}=m^\mu_{~~\mu}=0~~.
\ee
The relationship between $e$ and $\bre$ implies that
\be
\brh_{\mu\bra}=h_{\mu\bra}+m_{\mu\bra}+\phi\bre^{(0)}_{\mu \bra}~~,
\ee
or
\be
\brh_{\mu\nu}=h_{\mu\nu}+m_{\mu\nu}+\phi\eta_{\mu \nu}~~,
\ee
together with corresponding equations in other bases.

To lowest order
\be
R_{ab\mu\nu}=\d_\mu\om_{\nu ab}-\d_\nu\om_{\mu ab}~~.
\ee
Hence
\be
R^{\ss}_{~~\mu}=e^{(0)a\ss}e^{(0)b\nu}R_{ab\mu\nu}~~,
\ee
so that
\be
R^{\ss}_{~~\mu}=\d_\mu\om_{\nu}^{~~\ss\nu}-\d_{\nu}\om_\mu^{~~\ss\nu}~~,
\ee
and
\be
R=2\d_\mu\om_\nu^{~~\mu\nu}~~.
\ee
Again in the lowest order approximation
\be
j_{\mu ab}=\d_\mu m_{ab}+\om_{\mu ab}-\bom_{\mu ab}~~.
\ee
If we convert to the coordinate basis we have
\be
j_{\mu\ll\tau}=\d_\mu m_{\ll\tau}+\om_{\mu\ll\tau}-\bom_{\mu\ll\tau}~~.
\ee
We can also evaluate $\brj_\mu$~. In this lowest approximation it
coincides with $j_\mu$~. From eq(\ref{EQM3}) we have
\be
\brj_{\mu[\ll,\tau]}=j_{\mu[\ll,\tau]}=0~.
\ee
That is
\be
\d_\ss m_{[\ll,\tau]}+\om_{\ss\ll\tau}-\bom_{\ss\ll\tau}=0~~.
\label{CON4}
\ee
From eq(\ref{EQM3}) we see that the torsion in the gravitational vierbein bundle vanishes.
Explicitly we have
\be
C^\ll_{\mu\nu}=\fr{1}{2}\left(\d_\mu h_\nu^{~~\ll}-\d_\nu h_\mu^{~~\ll}
        +\om_{\mu~~\nu}^{~\ll}-\om_{\nu~~\mu}^{~\ll}\right)=0~~.
\label{CON5}
\ee

Eq(\ref{EQM2}) yields
\be
\eta^{\mu\ss}\d_\mu\left(\d_\ss m_{\ll\tau}+\om_{\ss\ll\tau}
                        -\bom_{\ss\ll\tau}\right)+m^2m_{\{\ll,\tau\}}=-8\pi FU_{\ll\tau}~~,
\ee
where
\be
U_{\ll\tau}=\BT_{\ll\tau}-\fr{1}{4}\eta_{\ll\tau}\BT~~,
\ee
and we have assumed that $\BT_{\mu\nu}$ and hence $U_{\mu\nu}$ is a first order quantity.
Making use of eq(\ref{CON4}) we obtain the result
\be
\eta^{\mu\ss}\d_\mu\d_\ss m_{\left\{\ll,\tau\right\}}+m^2m_{\{\ll,\tau\}}
                 =-8\pi F\left(\BT_{\ll\tau}-\fr{1}{4}\eta_{\ll\tau}\BT\right)~~.
\ee
Because we set $\gamma=8$ in eq(\ref{EQM1}) we obtain
\be
R_{\ss\ll}-\fr{1}{2}\eta_{\ss\ll}R=8\pi G\BT_{\ss\ll}~~.
\ee

In the present approximation $m_{[\mu,\nu]}$ can be removed by gauge
transformations of the form
\be
\om_{\ss\ll\tau}\rightarrow \om_{\ss\ll\tau}+\d_\ss\phi_{\ll\tau}~~,
~~~~~~~~\mbox{and}
~~~~~~~~\bom_{\ss\ll\tau}\rightarrow \bom_{\ss\ll\tau}+\d_\ss{\bar \phi}_{\ll\tau}~~.
\ee
We can assume therefore that in this approximation $m_{[\mu,\nu]}$ vanishes.
Therefore $m_{\mu\nu}$ may be assumed symmetric. It satisfies
\be
(\d^2+m^2)m_{\mu\nu}=-8\pi F(\BT_{\mu\nu}-\fr{1}{4}\eta_{\mu\nu}\BT)~~.
\label{WE1}
\ee
The gauge invariance referred to above means also the we are free to
choose $h_{\mu\nu}$ to be symmetric with the result that $\brh_{\mu\nu}$
is also symmetric.
Under these circumstances we can solve eq(\ref{CON5}) to yield
\be
\om_{\nu\ll\mu}=\d_\mu h_{\nu\ll}-\d_\ll h_{\nu\mu}~~.
\ee
Eq(\ref{EQM1}) now implies that
\be
\d_\mu\d_\nu h^\nu_{~~\ss}+\d_\ss\d_\nu h^\nu_{~~\mu}-\d^2h_{\mu\ss}
-\d_\mu\d_\ss h^\nu_{~~\nu}-\eta_{\ss\mu}(\d_\nu\d_\tau h^{\nu\tau}-\d^2h^\nu_{~~\nu})
=8\pi G\BT_{\ss\mu}~~.
\ee
We now refine our coordinate system by choosing the harmonic gauge.
\be
g^{\mu\nu}\GG^\ll_{\mu\nu}=0~~.
\ee
In the lowest approximation it yields
\be
\d_\mu h^\mu_{~~\ll}=\fr{1}{2}\d_\ll h^\mu_{~~\mu}~~.
\label{GC}
\ee
The equation of motion then becomes
\be
\d^2\left(h_{\mu\ss}-\fr{1}{2}\eta_{\mu\ss}h^\tau_{~~\tau}\right)=-8\pi G\BT_{\mu\ss}~~,
\ee
or
\be
\d^2 h_{\mu\ss}=-8\pi G\left(\BT_{\mu\ss}-\fr{1}{2}\eta_{\mu\ss}\BT\right)~~.
\label{WE2}
\ee
The equation for $\phi$ is
\be
(\d^2+m^2)\phi=-8\pi F'\BT~~.
\ee

To demonstrate how the theory matches up to the standard tests of General
Relativity we consider its implications for phenomena on a 
scale such as the solar system that is much smaller than the galactic scale. 
For such applications we can set $m=0$~. We will return to the problem with 
the galactic scale parameter later.

By combining the above equations we find for the massless case
\be
\d^2 \brh_{\mu\nu}=-8\pi(G+F)\BT_{\mu\nu}+8\pi\left(\fr{1}{2}G
                                          +\fr{1}{4}F-F'\right)\eta_{\mu\nu}\BT~~.
\ee

By making use of the result $\d_\ss\BT^\ss_\nu=0$, we can show also
that
\be
\d^2\left[\d_\mu\brh^\mu_{~~\nu}-\xi\d_\nu\brh^\mu_{~~\mu}\right]=0~~,
\ee
where
\be
\xi=\fr{\fr{1}{2}G+\fr{1}{4}F-F'}{G-4F'}~~.
\ee
Assuming then that the relevant fields are sourced by the energy 
momentum tensor we can maintain the condition
\be
\d_\mu\brh^\mu_{~~\nu}-\xi\d_\nu\brh^\mu_{~~\mu}=0~.
\ee
For arbitrary $F'$ this gauge condition on $\brh_{\mu\nu}$ is different
from that on $h_{\mu\nu}$~. However the choice 
\be
F'=-\fr{1}{4}F~~,
\ee
is of special interest. We have
then $\xi=\fr{1}{2}$~. As a result $\brh_{\mu\nu}$ satisfies the same harmonic
condition as $h_{\mu\nu}$~. 
\be
\d_\mu\brh^\mu_{~~\nu}-\fr{1}{2}\d_\nu\brh^\mu_{~~\mu}=0~.
\label{GCBAR}
\ee
The wave equation for $h_{\mu\nu}$ also takes a significant
form and becomes
\be
\d^2\brh_{\mu\nu}=-8\pi(G+F)\left(\BT_{\mu\nu}-\fr{1}{2}\eta_{\mu\nu}\BT\right)~~.
\label{WEBAR}
\ee
It follows that for this choice of $F'$ weak field gravity is related to
the matter distribution exactly as in General Relativity if we set
Newton's constant $G_N=G+F$~. Hereafter we will assume that $F'$ has this
special value and that Newton's constant is obtained in this way from the theory.

For example in a static situation where $\d^2=-\nabla^2$, $\BT_{00}=\BT=\rho$
(the density of matter)  we have 
\be
\nabla^2\brh_{00}=4\pi G_N\rho~~.
\ee
Of course we can interpret $\brh_{00}$ as the gravitational
potential experienced by a material particle. In addition the spatial part of the metric
satisfies
\be
\nabla^2\brh_{ij}=4\pi G_N\rho\dd_{ij}~~.
\ee
This is precisely the form for the spatial metric to yield Einstein's
prediction for the deflection of light and to satisfy the time-of-flight measurements
of radio signals \cite{CM1,CM2}~.

The remaining solar system scale test is the precession of the orbit of Mercury.
This requires a higher order correction than the Newtonian approximation of the  weak 
field limit. For reasons of space we do not present this calculation here but
we have checked the consistency of the theory on this point by examining the
asymptotic behaviour of the Schwarzschild-like solution.  The result is that
the refinement of the Newtonian potential that produces the precession is correctly
given by the theory.

It is clear from the wave equation for $\brh_{\mu\nu}$, eq(\ref{WEBAR}), and the associated gauge 
condition, eq(\ref{GCBAR}), that the gravitational waves emitted by a time dependent matter 
distribution will be exactly the same as predicted by General Relativity. The detection
of these waves by ordinary matter will also be entirely conventional. It is reasonable
to conclude that the observations of the slowing of a binary quasar and its conformity 
with the predictions of GR will be reproduced in our theory \cite{HULTAY,HUL,TAY}.

There are circumstances in which the bi-metric theory could show differences with
General Relativity. These would occur were there to be a form of matter that coupled directly
to the metric $g_{\mu\nu}$~. Such matter would act as a source for and react to 
the unbarred metric. In the weak field limit it would therefore behave as if
$G_N=G$ and if mixed with other matter would imply that the Equivalence Principle
did not hold. Part of our theory of ordinary matter and gravity is that such
anomalous matter is not present. This is the assumption we use throughout this paper.

\section{``Dark Matter'' and the Galactic Scale}

In order to apply our theory to objects of galactic or extra-galactic size
we restore the ``mass'' parameter $m$~. Our hypothesis is that $m^{-1}$
is a length of the order of $30$ kpc. Assigning such a value to $m^{-1}$ 
achieves the purpose that the theory exhibits three regions of length scale, namely
(i) 0-1 kpc in which conventional Newtonian gravity holds sway and for which the theory 
of the previous section is relevant, (ii) 1-100 kpc, a transition region
appropriate to galactic dynamics and (iii) above 100 kpc in which Newtonian gravity
with an enhanced coupling appears.

To analyse the gravitational effect of the galactic scale we set 
\be
\brh_{\mu\nu}=h_{\mu\nu}+h'_{\mu\nu}~~,
\ee
where
\be
h'_{\mu\nu}=m_{\mu\nu}+\phi\eta_{\mu\nu}~~,
\ee
while $h_{\mu\nu}$ satisfies eq(\ref{WE2}) $h'_{\mu\nu}$ satisfies
\be
(\d^2+m^2)h'_{\mu\nu}=-8\pi F\left(\BT_{\mu\nu}-\fr{1}{2}\eta_{\mu\nu}\BT\right)~~.
\ee
Therefore while a highly localised matter distribution of mass M, at the origin yields
\be
h_{00}=-GM\fr{1}{r}~~
\ee
it gives rise to 
\be
h'_{00}=-FM\fr{e^{-mr}}{r}~~,
\ee
for the remaining part of the metric. We have then
\be
\brh_{00}=-\fr{GM}{r}\left(1+\fr{F}{G}e^{-mr}\right)~~.
\ee
It follows that the effective gravitational coupling for $mr<<1$ is $G_N=G+F$
while at distances beyond the galactic scale for which $mr>>1$, the effective coupling 
is $G$~. Our hypothesis is that $G_N<G$ so that gravity is weaker at short distances
than at long distances. This is the basis of our explanation for ``dark matter''.
For this to be true we clearly must have $F<0$~. We set
\be
F=-\eps G~~,
\ee
with $\eps>0$, so the above gravitational potential becomes
\be
\brh_{00}=-\fr{G_N M}{(1-\eps)}\fr{\left(1-\eps e^{-mr}\right)}{r}~~.
\ee
More generally for a matter distribution $\rho(\r)$ we have
\be
\brh_{00}(\r)=-\int d^3\r'\rho(\r')\fr{G_N}{(1-\eps)}\fr{(1-\eps e^{-m|\r-\r'|})}{|\r-\r'|}~~.
\ee
We believe that this choice of a negative sign for $F$ is in fact the ``natural'' choice
and stabilises the theory against tumbling of coordinate frames \cite{CW1}~. We will pursue 
this analysis in a future publication. Here we explore the more immediate physical implications 
of our assumptions.

\subsection{Galactic Clusters}

For objects much greater than galactic size such as galactic clusters,
there is ample evidence that the observed matter is insufficient to account for
the gravitational potential inferred from applications of the virial theorem
to the motion of galaxies in the cluster \cite{ZWI1, CAR}.
Currently the picture of such clusters is that 5-10\% of the mass is galactic in origin
with a further contribution from hot gas. The bulk of the gravitational potential
is accounted for by dark matter. It is also significant that a detailed analysis of 
clusters suggests that the dark matter distribution follows that of the visible
matter \cite{CAR}. The clusters appear to be condensed versions of the local background.

The explanation of this effect is straightforward in our model. The appearance of 
dark matter is simply the consequence of an enhanced gravitational coupling of visible 
matter. The parameter $\eps$ represents the dark fraction of apparent matter.
On the basis of the above observations we should expect $\eps\simeq 0.9-0.95$~.
Note that because our extragalactic dynamics is still Newtonian, although with an enhanced
coupling, it is still possible to apply analyses of galactic clusters that rely on the  
virial theorem for an inverse square law of force between galaxies \cite{ZWI1}.

The inverse square law is also crucial for the argument relating the motion of the Local  
Group to optical flux due to galaxies \cite{LYN}.  The close alignment of the motion of the 
Local Group through the CMB with the net optical flux confirms that fluctuations in the 
distribution of visible and dark matter are closely correlated.  Disparities in the 
visible and dark matter distributions would tend to destroy this alignment. In our 
theory of course, the distribution of dark matter is identical to that of visible matter 
of which it is merely a reflection.  There is no room for any bias between visible 
and dark matter fluctuations.

We will see later that the re-interpretation of dark matter as an enhancement 
of the Newtonian constant at large scales extends also to the dynamics of the
expanding universe.

\subsection{Galactic Rotation Curves}

On a galactic scale we are concerned with the rotation curves of galaxies. The apparent 
asymptotic flatness of many of these curves is usually taken as the most direct evidence 
for dark matter \cite{RUB1}-\cite{ALB1}.  Our explanation rests on the modification
of the gravitational interaction described above. It turns out that by choosing
a value for $\eps$ in the range suggested by large scale dynamics it is possible, using
a disk model, appropriate for certain spiral galaxies, to compute rotation curves that 
exhibit the features of observed rotation curves. 

As an example we construct a model galaxy with a thin disk of surface density
\be
\ss(r)=\ss_de^{-a_dr}~~.
\ee
The mass of the disk is $m_d=2\pi\ss_d/a_d^2$~. The gravitational potential in the plane
of the galaxy is
\be
\psi(r)=-\fr{G_N\ss_d}{1-\eps}
        \int d^2\r'e^{-a_dr'}\fr{1}{|\r-\r'|}\left(1-\eps e^{-m|\r-\r'|}\right)~~,
\ee
and the tangential rotational velocity, $v$, is given by
\be
v^2=\r.\nabla\psi(r)=\fr{G_N\ss_da_d}{1-\eps}
 \int d^2\r'e^{-a_dr'}\r.{\hat\r'}\fr{1}{|\r-\r'|}\left(1-\eps e^{-m|\r-\r'|}\right)~~.
\ee

The galaxy NGC 3198 has a well measured rotation curve and it is accepted that a simple 
exponential disk model gives a good account of its luminosity distribution \cite{ALB1}. 
We assume a mass distribution with the same exponential shape. In Fig \ref{NGC3198} we show the 
resulting rotation curve where we have chosen the galaxy parameters to be 
$a_d=0.38 \mbox{kpc}^{-1}$ and $m_d=2.9\times 10^{10}M_\odot$, these are quite close to 
the values of a previous analysis \cite{ALB1}. The theory parameters are chosen to be 
$\eps=0.937$ and $m=0.035~\mbox{kpc}^{-1}$ or $m^{-1}=28.6$ kpc. We stress 
that these parameters are not a best fit but merely the result of eyeball exploration.
However it is not easy to reduce the value of $\eps$ value by much and obtain a convincing 
shape.  It is encouraging that $\eps$ does lie in the range we anticipated from our 
discussion of extragalactic structure.

The Fig \ref{NGC3198} also shows the standard Newtonian curve for the assumed mass 
distribution. The difference between the two curves is normally attributed to dark matter.
Here we achieve the same effect by means of our modified gravity theory. Fig \ref{NGC3198L} shows
the extrapolation of the curve to larger radius. There is clearly a beginning of a fall-off 
around 90-100 kpc although the there is still substantial rotational velocity out to 400 kpc.
The rotation curve becomes Newtonian but with an enhanced gravitational coupling
\be
v\simeq\sqrt{\fr{GM_d}{r}}~~.
\ee

Of course the challenge to our theory is to fit all galactic rotation curves simultaneously
with common values for $m$ and $\eps$~ and plausible masses and and mass distributions
for the galaxies.  This is no easy task since many galaxies have a more complicated structure 
than NGC3198 and detailed modelling will be required to determine the adequacy of our 
theoretical predicitions.  We intend to pursue this task in the future. However one 
prediction of the theory does not require such a detailed attack. We predict that 
{\it all} galactic rotation curves will fall away in the range 100-200 kpc from the 
galactic centre.  Measurements in this range and beyond  would be a direct test of our ideas.

Our proposal of a galactic scale determining the shapes of rotation curves is in conflict 
with the construction of a proposed universal scaling curve for galaxies  \cite{PER2}. 
This approach takes the optical radius of the galaxy as the significant length scale
and uses a statistical approach to establish well defined rotation curves. We 
are proposing a model in which the ratio of the optical radius to the galactic
scale is the significant quantity for establishing the shape of rotation curves. The statistical
approach deployed in \cite{PER2}, which superimposes results from galaxies of different sizes,
is therefore not open to us. It may well be however that the detailed modelling required 
to test our theory will provide an equally good description of actual galactic rotation 
curves. More importantly, if a common scale for the underlying structure of all galaxies were 
established it holds out the prospect of better distance measurements and hence the possibility 
of measuring the Hubble constant with more certainty.

\subsection{Expanding Universe}

Finally we wish to show that when the matter density is small as in the present epoch, 
the expansion of the universe is controlled by the standard equations of General Relativity
but with Newton's constant, $G_N$, replaced by the enhanced constant $G$~. In other words
the dark matter component required to relate the visible matter to
the observed Hubble constant is again supplied by the enhancement mechanism. Of course a
complete reconciliation of the equations with observation requires a contribution
to the energy density from the Cosmological Constant. This necessity can be catered for
also in our theory. 

To apply our theory we take the usual starting point that the spatial sections of the 
universe are isotropic and homogeneous. In the present approach we meet these requirements 
by assuming that there exist basis vierbein fields, $E_{\mu a}$ such that 
$E_{ta}E_{tb}\eta^{ab}=1$ and $E_{ta}E_{\mu b}\eta^{ab}=0$ for $\mu\ne t$~.  We choose our 
coordinates so that $E_{ta}$ is constant in space-time and $E_{\mu a}$ depends only on 
coordinates on the spatial section when $\mu\ne t$~. We finally fix the nature of the 
space-time model by choosing structure constants $f_{bca}$ so that
\be
\d_\mu E_{\nu a}-\d_\nu E_{\mu a}=-f_{bca}E_\mu^{~~b}E_\nu^{~~c}~~.
\ee
Of course $f_{bca}$ vanishes if any of the suffixes is 0 (timelike). For a spatially flat universe
all structure constants have the value zero. For a universe of positive
spatial curvature the spatial constants $f_{jki}=2\eps_{jki}$, where 2 is a convenient
normalization, and for a negatively curved universe $f_{jki}=\dd_{ij}n_k-\dd_{ik}n_j$ 
where $n_k$ is an arbitrary unit three-vector.

We construct the expanding universe by choosing the gravitational vierbein to have the form 
\be
e_{\mu a}=A_aE_{\mu a}~~,
\ee
where
\be
A_0=1~~~~~~~~\mbox{and}~~~~~~~~A_i=A(t)~~,
\ee
A similar structure is maintained for the matter vierbein by requiring the transformation
matrix $M$ to be diagonal. That is (we equivalence the labels $a$ and $\bra$ in the obvious way),
\be
\bre_{\mu\bra}=e^\phi \LL_ae_{\mu a}=\BA_aE_{\mu a}~~~~~~~~
                                           \mbox{and}~~~~~~~~\BA_a=\LL_aA_ae^{\phi}~~,
\ee
where $\LL_0=\LL(t)$ and  $\LL_i=\LL_S(t)$~. The requirement that $\det M=1$ implies that
$\LL_S=\LL^{-\fr{1}{3}}$~.

Using the above information and eq(\ref{VBCON1}) we find that
\be
\hat\om_{i0j}=-\hat\om_{ij0}=\fr{\dot A}{A}\dd_{ij}~~,
\ee
and
\be
\hat\om_{ijk}=-\fr{1}{2A}\left(f_{ijk}+f_{kij}+f_{kji}\right)~~.
\ee
All other components vanish. A similar structure is found for $\om_{abc}$~. From eq(\ref{VBCON2}) 
we obtain
\be
\om_{i0j}=-\om_{ij0}=\fr{\dot A}{AE}\dd_{ij}~~,
\ee
where
\be
E=1-\fr{G}{4F}\left(2-\fr{\LL^2}{\LL_S^2}-\fr{\LL_S^2}{\LL^2}\right)~~,
\ee
and
\be
\om_{ijk}=\hat\om_{ijk}~~.
\ee

From eq(\ref{CURV1}) we find for the relevant components of the curvature 
tensor
\be
R_{00}=-\fr{3}{A}\d_t\left(\fr{\dot A}{E}\right)~~,
\ee
and
\begin{eqnarray}
R_{ij}&=&\left(\fr{1}{A}\d_t\left(\fr{\dot A}{E}\right)
                    +2\left(\fr{\dot A}{AE}\right)^2\right)\dd_{ij}\\\nonumber
  &&-\fr{1}{4A^2}\left(f_{kjl}+f_{ljk}+f_{lkj}\right)\left(f_{lik}+f_{kil}+f_{kli}\right)\\\nonumber
  &&+\fr{1}{2A^2}\left(f_{jik}+f_{kji}+f_{kij}\right)f_{lkl}~~.
\end{eqnarray}
If we evaluate this expression for the three curvature cases we find
\be
R_{ij}=\left(\fr{1}{A}\d_t\left(\fr{\dot A}{E}\right)
                    +2\left(\fr{\dot A}{AE}\right)^2+\fr{2k}{A^2}\right)\dd_{ij}~~,
\ee
where conventionally $k=0,\pm 1$ according as the spatial curvature is zero, positive or negative.
We also have from eq(\ref{STRCUR})
\be
j_{tab}j_t^{~ba}=\fr{4}{3}\left(\fr{\dot \LL}{\LL}\right)^2~~,
\ee
and
\be
j_{iab}j_j^{~ba}=\fr{1}{2}\left(\fr{\dot A}{AE}\right)^2
                  \left(2-\fr{\LL^2}{\LL_S^2}-\fr{\LL_S^2}{\LL^2}\right)\dd_{ij}~~.
\ee

If we denote the matter energy density and pressure by $\rho$ and $p$ respectively,
then from eq(\ref{EQM1}) we obtain
\begin{eqnarray}
-\fr{3}{A}\d_t\left(\fr{\dot A}{E}\right)-\fr{4G}{3F}\left(\fr{\dot \LL}{\LL}\right)^2
                    -\fr{G}{F'}{\dot \phi}^2~~~~~~~~~~~~~~~~~~~~&&\\\nonumber
+\fr{G}{F}\fr{m^2}{8}\left(\LL^2+\LL^{-2}+3\LL_S^2+3\LL_S^{-2} -\gamma\right)
             +\fr{G}{F'}\fr{m^2}{2}\phi^2 &=&4\pi Ge^{4\phi}\left(\rho+3p\right)~~,
\label{NEQM1}
\end{eqnarray}
and
\begin{eqnarray}
\fr{1}{A}\d_t\left(\fr{\dot A}{E}\right)+2E\left(\fr{\dot A}{AE}\right)^2
                                     +\fr{2k}{A^2}~~~~~~~~~~~~~~~~~~~~~~&&\\\nonumber
-\fr{G}{F}\fr{m^2}{8}\left(\LL^2+\LL^{-2}+3\LL_S^2+3\LL_S^{-2}
                  -\gamma\right)-\fr{G}{F'}\fr{m^2}{2}\phi^2
     &=&4\pi Ge^{4\phi}(\rho-p)~~.
\label{NEQM2}
\end{eqnarray}
From eq(\ref{EQM2}) we find
\begin{eqnarray}
\d_t\left(\fr{\dot \LL}{\LL}\right)+3\fr{\dot A}{A}\fr{\dot\LL}{\LL}
-\fr{3}{2}\left(\fr{\dot A}{AE}\right)^2
                  \left(\fr{\LL^2}{\LL_S^2}-\fr{\LL_S^2}{\LL}\right)\\\nonumber
+\fr{3m^2}{16}\left(\LL^2-\LL^{-2}-\LL_S^2+\LL_S^{-2}\right)
                                    &=&-6\pi Fe^{4\phi}\left(\rho+p\right)~~.
\label{NEQM3}
\end{eqnarray}
Finally from eq(\ref{EQM5}) we obtain
\be
{\ddot\phi}+\fr{3{\dot A}}{A}{\dot\phi}+m^2\phi=-8\pi F'e^{4\phi}\left(\rho-3p\right)~~.
\label{NEQM4}
\ee

Eq(\ref{NEQM1}) and eq(\ref{NEQM2}) can be combined to eliminate the second derivative in $t$~.
\begin{eqnarray}
E\left(\fr{\dot A}{AE}\right)^2+\fr{k}{A^2}-\fr{2}{9}\fr{G}{F}\left(\fr{\dot\LL}{\LL}\right)^2
      -\fr{1}{6}\fr{G}{F'}{\dot\phi}^2~~~~~~~~~~~~~~~~~~~~~~~~~~~~~~&&\\\nonumber
-\fr{G}{F}\fr{m^2}{24}\left(\LL^2+\LL^{-2}+3\LL_S^2+3\LL_S^{-2}-\gg\right)
        -\fr{G}{F'}\fr{m^2}{6}\phi^2&=&\fr{8\pi G}{3}e^{4\phi}\rho~~.
\label{NEQM5}
\end{eqnarray}
If we multiply this equation by $A^2$ and differentiate with repect to $t$
we can use the above equations of motion to deduce that
\be
{\dot\rho}=\left(\fr{\dot\LL}{\LL}-3\fr{\dot A}{A}-3{\dot\phi}\right)(\rho+p)~~.
\ee
This is easily re-expressed as
\be
\d_t\left(\rho\fr{e^{3\phi}A^3}{\LL}\right)+p\d_t\left(\fr{e^{3\phi}A^3}{\LL}\right)~~,
\ee
or
\be
\d_t\left(\rho\BA^3\right)+p\d_t\left(\BA^3\right)~~,
\label{CONSV}
\ee
where $\BA=e^\phi A\LL^{-\fr{1}{3}}$ is the cosmic radius parameter appropriate to matter. 
Eq(\ref{CONSV}) is therefore the standard equation for the conservation of the energy-momentum
tensor in this special case, as we should have expected. 

The above equations are rather complicated but can be simplified if we ask how
they might be applied to our expanding universe in which the pressure vanishes and 
the matter density is low and getting lower. In the absence of matter eq(\ref{NEQM5}) 
has a solution for which $\phi\equiv 0$~. For weak density we assume that there is 
a solution for which $\phi\simeq O(\rho)$~. We have then as a leading approximation
\be
{\ddot\phi}+\fr{3{\dot A}}{A}{\dot\phi}+m^2\phi=-8\pi F'\rho~~.
\ee
The form of this equation suggests that, ignoring transients, the solution is
to a good approximation
\be
\phi=-\fr{8\pi F'}{m^2}\rho=-\fr{2\pi \eps G}{m^2}\rho~~.
\ee
We expect then that $\phi$ will remain $O(\rho)$~. If we set $\LL=e^\xi$ and
omit all terms $O(\xi^2)$ then $E\simeq 1$ and eq(\ref{NEQM4}) becomes
\be
\ddot\xi+3\fr{\dot A}{A}{\dot\xi}
    +\left(m^2-8\left(\fr{\dot A}{A}\right)^2\right)\xi=-6\pi F\rho~~.
\ee
Again we expect there to be a solution $\xi=O(\rho)$~. 
If now we neglect all terms $O(\rho^2)$ we obtain the equations
\be
\left(\fr{\dot A}{A}\right)^2+\fr{k}{A^2}=\fr{8\pi G}{3}(\rho+\rho_{CC})~~,
\ee
where the energy density associated with the cosmological constant, $\rho_{CC}$
is given by
\be
\rho_{CC}=\fr{m^2}{8\pi F}\left(1-\fr{\gg}{8}\right)
                  =\fr{m^2(1-\eps)}{8\pi G_N\eps}\left(\fr{\gg}{8}-1\right)~~.
\ee
and
\be
\d_t\left(\rho A^3\right)=0~~.
\ee
These are the standard equations for the expanding universe but with
Newton's constant $G_N$ replaced by the enhanced constant $G$~. To lowest order
in $\rho$ the parameter $t$ is also the proper time of comoving matter in its
own metric since the two metrics coincide under these circumstances. It
follows that Hubble's constant $H_0={\dot A/A}$ and the current deceleration
parameter is $q_0=-{\ddot A}A/{\dot A}^2$ and that the critical density
for a flat universe is
\be
\rho_c=\fr{3H_0^2}{8\pi G}=\fr{3H_0^2}{8\pi G_N}(1-\eps)~~.
\ee
So $\rho_c$ in our theory is between one tenth and one twentieth the value appropriate to standard 
General Relativity. This brings it within range of visible matter \cite{TYT}. In fact recent 
measurements of distant supernovae \cite{PERL} lead to an estimate of $q_0$ that suggests
$\rho=0.3\times \rho_c$ and $\rho_{CC}=0.7\times \rho_c$ (assuming a flat universe \cite{BER})
with the further implication that in the present theory, visible matter may support the current 
expansion of the universe albeit with the help of a cosmological constant.

\section{Conclusions}

In this paper we have constructed a modified theory of gravity that fits
all the standard tests of General Relativity that can be made on the scale of 
the solar system including the bending of light by the sun, time delay measurements
and the precession of Mercury's orbit. The theory is a bi-metric theory of a novel
kind with a very geometrical structure. It has the flexibility to permit
the introduction of a galactic length-scale of roughly 30 kpc. Gravitational
effects of matter at distances below the galactic scale, in the solar system
for example, are of a conventional kind and of a strength determined by Newton's
constant, $G_N$~. This outcome is achieved in the theory as the result of a
competition between two gravitational effects, a strong attraction and a repulsion that is 
nearly as strong. Over the range of the galactic scale the theory allows the
repulsion to fade out exponentially leaving the much stronger underlying
gravity to show through. The gravitational effects of matter on this large scale
are also conventional in character but of a strength determined by the
much stronger underlying gravitational constant $G\simeq 10-20\times G_N$~.
These results are consistent with observations of galactic clusters and also apply to the
expansion of the universe as a whole. As a result the critical density for a flat
universe is between one tenth and one twentieth of the value calculated in General
Relativity. This makes it more plausible that visible baryonic matter can support the
expansion of the universe without dark matter, though supplemented by a cosmological
constant.

The theory provides a natural 
explanation for the tendency of dark matter to follow the distribution of visible matter
since the former is simply an amplification of the latter. Dark matter does not have a 
separate dynamics of its own. On this basis the apparent dark matter is not in any
way biased in its fluctuation structure relative to visible matter. 

We showed that at least for the galaxy NGC3198 the theory can reproduce
the rotation curve out to 30 kpc (the end of the measured range) equally as well
as dark matter models. The theory also predicts that the rotation curve
will fall away in the 100-200 kpc range and eventually tend to a standard
inverse square root of distance behaviour but with an apparent mass $10-20$
times the expected mass of the galaxy ($\sim 3\times 10^{10}M_{\odot}$ for NGC3198) if the
conventional Newtonian constant is used for the estimate. Our theory
however assigns the expected mass but uses the enhanced value for the
gravitational constant appropriate at extra-galactic distances. This
phenomenon should appear for all galaxies. It is therefore important
for the theory to test rotation curves at distances of $100-200$ kpc from
the galactic centre. It is also important to test the effectiveness of the
theory in a range of galaxies and to arrive at a consistent picture with
a common set of gravitational parameters. This task requires a good understanding 
of the structure of individual galaxies and and accurate knowledge of their distance.
However we note that the establishment of a fixed scale associated with all galaxies
would be of considerable help in establishing galactic distances.

In addition to the incomplete phenomenological analyses discussed above
there are many issues yet to be explored in our bi-metric theory. 
Of particular importance are the development of density fluctations, dynamics 
and formation of galaxies using the modified gravitational law, and evolution 
of the early universe.  In this last context the ability of bi-metric theories to support 
anomalous propagation of signals (VSL phenomena) will be of great importance. The 
existence within the theory of black holes is also a topic that it would be very interesting 
to resolve. On intuitive grounds it seems reasonable to suppose that black holes  will 
exist in our theory but the nature of the horizon may be more complicated than in General 
Relativity because of its double light-cone structure.

\section*{Acknowledgements}
I thank Jonathon Evans for helpful and clarifying discussions.
\pagebreak

\newpage
\befh
\bec
\epsfig{file=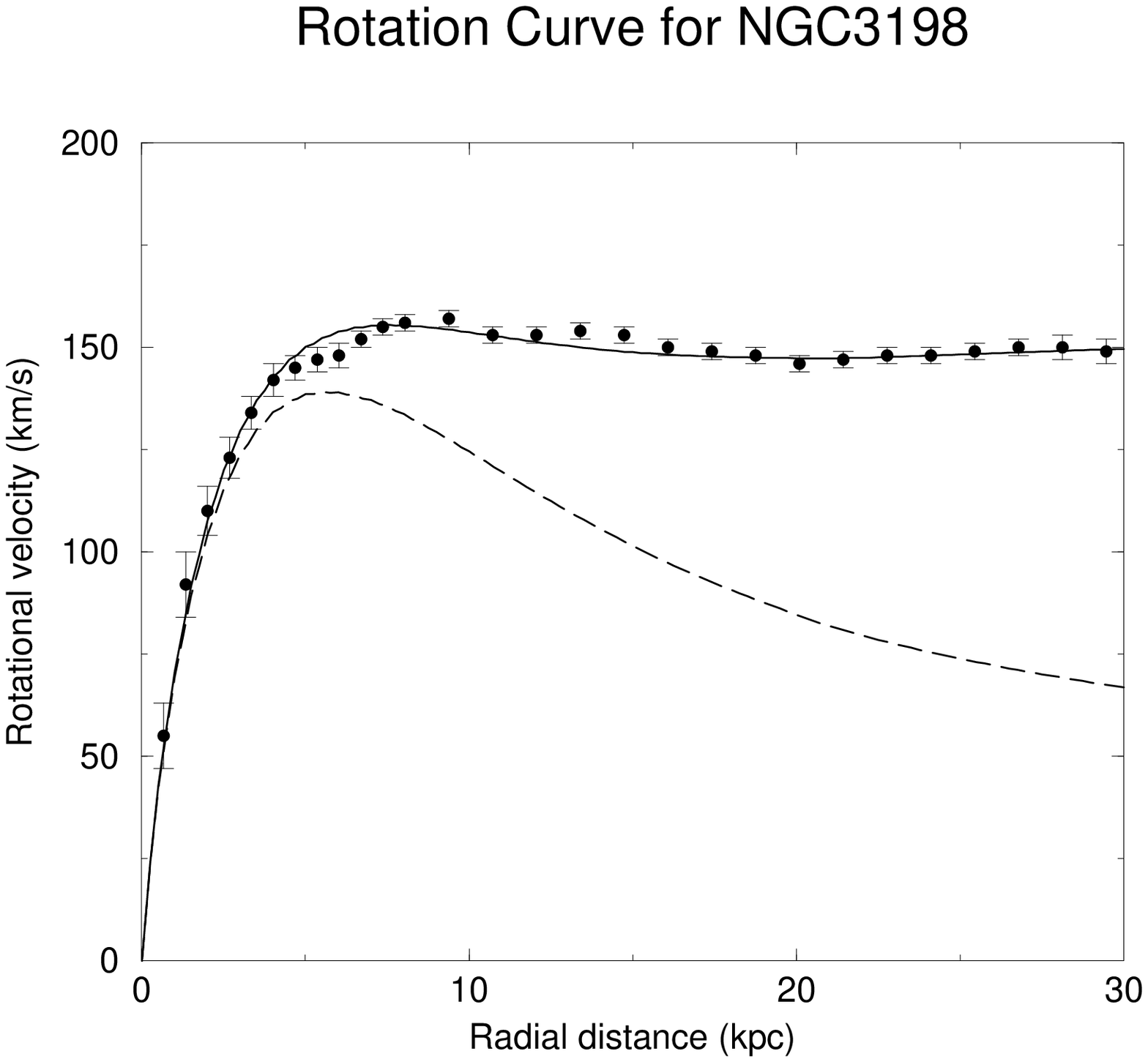,height=120mm}
\enc
\caption{\label{NGC3198}\small
Data for NGC3198 compared to exponential disk model with enhanced
gravity (full line) and the Newtonian (no enhancement) curve for 
the same mass (dashed line).
}
\enf

\befh
\bec
\epsfig{file=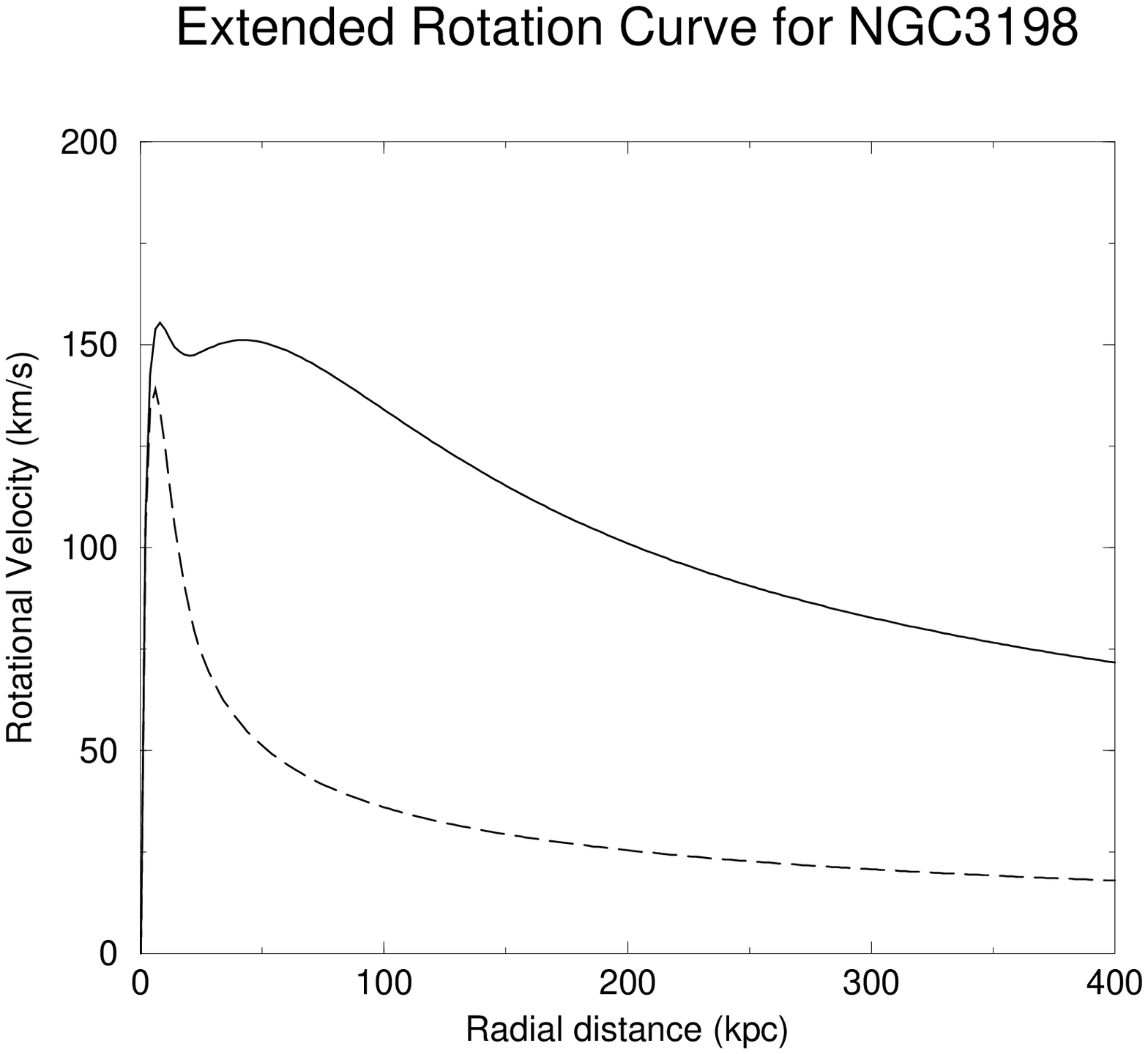,height=120mm}
\enc
\caption{\label{NGC3198L}\small
The extended rotation curve for the exponential disk model with enhanced
gravity (full line) and the Newtonian (no enhancement) curve for
the same mass (dashed line)
}
\enf

\end{document}